\documentstyle[titlepage,12pt]{article}
\begin{document}
\title {Exponential-Potential Scalar Field Universes II: The Inhomogeneous
Models }
\author { J. M. Aguirregabiria, A. Feinstein and J. Ib\'a\~nez\\
Dept.  de F\'{\i}sica Te\'orica, Universidad del Pa\'{\i}s Vasco,
Bilbao, Spain.}
\maketitle

\begin{abstract}
We obtain exact solutions for the Einstein equations with an
exponential-potential scalar field (\(V=\Lambda e^{k\phi}\)) which represent
simple
inhomogeneous generalizations of Bianchi I cosmologies. Studying these
equations
numerically we find that in most of the cases there is a certain period of
inflationary behaviour for \(k^2<2\). We as well find that for \(k^2>2\) the
solutions homogenize generically at late times. Yet, {\em none of the
solutions}
isotropize. For some particular values of the integration constants we find a
multiple
inflationary behaviour for which the deceleration and the inflationary phases
interchange
each other several times during the history of the model.

\end{abstract}

\section{Introduction}

In our previous paper \cite{AFI} (hereafter paper I) we have obtained a general
exact solution describing anisotropic Bianchi type I universes filled with
exponential-potential scalar field and studied their behaviour. These studies
are relevant in order to clarify as to how sensitive inflationary cosmologies
are to the pre-inflationary epoch characterized by different initial
conditions. For the case of Bianchi I models we confirm
previous results based on numerical, approximate and qualitative techniques
obtained by
various authors \cite{SA} predicting  the power law
inflationary  behaviour for a wide range of initial conditions.

Some of our latest studies \cite{fe},J\cite{prl} with more complicated
geometries,
however, cause a suspicion that the inflationary phenomena is not that generic
and
probably requires some special initial conditions. The Bianchi I cosmological
models,
studied in our previous paper, are too simple to derive definite conclusions
related to this question. Moreover, if one  assumes \cite{hz}  the existence
of the gravitational radiation background one may have  to allow for large
amplitude  inhomogeneities during the early stages of the universe. In such
scenarios the influence of gravitational waves on the evolution of the Universe
increases as one goes back in time. Once the wavelength of the gravitational
radiation is comparable to the size of the horizon  these
waves may not be described as homogeneous relativistic fluid but rather
must be seen as large amplitude shear and rotation inhomogeneities. This, and
the
desire to clarify as to how generic is inflation, leads one to consider more
general inhomogeneous spacetimes.

Another issue of interest, if the models inflate, is whether they
approach to a homogeneous and isotropic regime at late times. Some latest
qualitative  studies \cite{h}  as well as those based on exact anisotropic
scalar field solutions \cite{fe} show that there are difficulties with
isotropization in some model Bianchi universes. It is quite possible that the
inflation is not that effective device when isotropization is considered.
Simple
models considered previously using gravitational pulses-solitons might be
competing or complementary in order to relsolve this question. In these models
solitons \cite{ver}, \cite{cha} act as flattenig devices in inhomogeneous
cosmological models  which start
highly irregularly but approach Bianchi I universes filled with gravitational
radiation in a finite time. The Bianchi I universes in  turn are known to
isotropize quite easily.

Some numerical and qualitative work about the existence of inflationary phases
in
inhomogeneous cosmologies have been done previously. Kurki-Suonio et
al. \cite{K} have shown that in some cases the inhomogeneities may prevent the
universe from ever entering the inflationary stage. Goldwirth and Piran
\cite{GP} have studied both new and chaotic inflation  concluding that the
duration of inflation may be significantly reduced by the inhomogeneity, while
Calzetta and Sakellariadou \cite{CS} have looked at inhomogeneous, but
asymptotically FRW models, and have concluded that the Cauchy data must be
homogeneous over several horizons lenghts in order for inflation to occur.
These
works  point in the direction that sufficiently
irregular initial data  may cause problems for inflation.

Recently  two of us have found an exact solution describing
inhomogeneous exponential-potential scalar field cosmological model \cite{prl}.
The
behaviour of the solution was persistently non inflationary. Since this is
only a particular solution of the Einstein field equations it is
interesting to see wether the behaviour depicted by the model is shared by
a larger class of inhomogeneous cosmological solutions, or the solution
we have found is quite untypical.

In this paper we  restrict ourselves to study the effects of one
dimensional inhomogeneities on the evolution of the exponential potential
scalar field cosmologies. These inhomogeneities may be induced either by the
irregularities of the scalar field itself or  by the initial
inhomogeneities in the geometry due, for example, to the presence of
primordial gravitational waves as mentioned above.

We will first treat exactly the coupled Einstein-Klein-Gordon equations
reducing them to a single non linear ordinary differential equation
similarly to that discussed in paper I. In contrast to the
Bianchi I case we are not able to find the general solution to this
equation. Still, we can resolve it in some particular cases. These
particular exact solutions then serve us as a test for the numerical
analysis.

In the following Section II we present the Einstein field equations
and the way to solve them. In Section III several particular
solutions are given and discussed. The Section IV is devoted to
the qualitative and numerical analysis. We conclude and discuss our results in
the Section V.

\section{The Einstein Equations}

We will concentrate on solutions with one dimensional inhomogeneity. These can
be described by the generalized Einstein-Rosen space-times which admit an
Abelian Group of isometries \(G_2\) and include the Bianchi models of type
I-VII as particular cases and, therefore, the flat and open FRW solutions. The
line element is
\begin{equation}
ds^2=e^f(-dt^2+dz^2)+g_{ab}\,dx^a dx^b \qquad a,b=1,2\, .
\end{equation}
The functions \(f\) and \(g_{ab}\) depend on \(t\) and \(z\).

Assuming that
the two Killing vectors are hypersurface orthogonal  the line
element (1) may be cast in the following diagonal form
\begin{equation}
ds^2=e^f(-dt^2+dz^2)+G(e^p dx^2+e^{-p} dy^2)
\end{equation}
It is now apparent that (2) is a straitforward generalization of the models
considered in paper I but, here, the metric functions are allowed to depend on
\(t\)
and \(z\) variables.

To simplify the equations we shall only consider in this paper the class of
solutions for which the element of the transivity surface is homogeneous
\begin{equation}
G=G(t).
\end{equation}
This choice assures that the gradient of the transitivity surface area is
globally timelike \((G_{\mu}G^{\mu}<0)\) and hence is appropiate  for a
description of cosmological models \cite{Carm}.

The matter source for the metric is that of a minimally coupled scalar field
with   potential \(V(\phi)\) for which the stress-energy tensor is given by
\begin{equation}
T_{\alpha \beta}=\phi_{,\alpha}\phi_{,\beta}-g_{\alpha \beta}
\left(\frac{1}{2}\phi_{,\gamma}\phi^{,\gamma}+V(\phi) \right)\,.
\end{equation}
As in the paper I the potential is taken as
\begin{equation}
V(\phi) = \Lambda \, e^{k \phi}.
\end{equation}
One may rewrite this stress-energy tensor in a perfect fluid form (as long
as the gradient of the hypersurface \(\phi=constant\) is timelike) with the
kinematical and dynamical quantities of the fluid given in paper I.

For the line-element given by the Eq.(2) and the matter described by the
stress-energy tensor (4), the Einstein Equations can be written in the
following form:
\begin{equation}
\frac{\ddot G}{G} = 2\, e^{f}\, V\,,
\end{equation}
\begin{equation}
\ddot{p} -p''+ \frac{\dot G}{G}\, \dot{p} = 0 \,,
\end{equation}
\begin{equation}
\frac{1}{2}\dot{p}\, p'+\dot{\phi}\,\phi'-\frac{1}{2}\,f'\,\frac{\dot{G}}{G}=0
\end{equation}
\begin{equation}
\frac{\ddot G}{G} - \frac{1}{2}\left(\frac{\dot G}{G}\right)^2 - \frac{\dot
G}{G}
\,\dot{f} + \frac{ 1}{2}\,\dot p^2 +\frac{1}{2}\,p'^2 +
\dot{\phi}^2+\phi'^2=0\,.
\end{equation}
The Klein-Gordon equation for the scalar field is
\begin{equation}
\ddot{\phi}-\phi''+ \frac{\dot{G}}{G}\,\dot{\phi} +e^{f}\,
\frac{\partial{V}}{\partial{\phi}} = 0 \,.
\end{equation}

Without any loss of generality we write the scalar field as:
\begin{equation}
\phi=-\frac{k}{2} \log G + \Phi(t,z).
\end{equation}
Substituting the Eq.(11) into the Eq.(10) and using the form of the potential
given by the Eq.(5) along with the Eq.(6) we get the following equation for the
function \(\Phi\)
\begin{equation}
\ddot{\Phi}-\Phi''+ \frac{\dot{G}}{G}\,\dot{\Phi} = 0 \,.
\end{equation}
Note  that again, as in the case of the Bianchi I models the scalar field
\(\Phi\) and the transversal gravitational degree of freedom \(p\) verify the
same differential equation. This property is quite surprising and holds
apparently only in two cases: i) when the scalar field is massless  and
ii) when the scalar field has an exponential potential.

The case i) was studied
thoroughly by several authors in connection with quantum description of the
matter fields in an anisotropic background of the early universe \cite{Fe}.
By identifying the scalar field with the velocity potential of the
irrotational stiff fluid \cite{Ze}, Liang as well as Carmeli et al. \cite{C}
analysed within a fully non linear relativistic approach the development of
inhomogeneities on the spatially homogeneous background.

For the scalar field
with exponential potential this observation of the similarity between the two
equations is new and is helpful not only to construct exact solutions but to
see
the effects in separation of each of the fields on the dynamics of the models.
Not only these fields follow the similar equations but contribute equally to
the inhomogeneity as we shall see later.

We now suppose one may separate the functions \(p\) and \(\Phi\) in the
following way
\begin{equation}
p=\Pi(t)+P(z), \quad \Phi=\chi(t)+\psi(z).
\end{equation}
One may separate the solutions yet in a different way as products, however, one
may
prove that this only leads to a particular case of the solutions obtained by
the separation
(13). This basically happens because of the restrictive conditions imposed by
the Eq.(8)
and the form of the scalar field potential.

Substituting the expressions (13) into the Eqs.(7)
and (12) one obtains that \begin{equation} p=\Pi(t)+\frac{1}{2} \lambda\,
z^2+\gamma\, z,
\quad \Phi=\chi(t)+\frac{1}{2} l\, z^2+g\, z \end{equation}
where \(\lambda,J\gamma, l\) and \(g\) are constants.

Substituting these equations into the Eq.(8) and using \(f'\) obtained on
differentiating the Eq.(6) one gets the following condition
\begin{equation}
\frac{1}{2}\, \dot\Pi\, (\lambda z+\gamma)+\dot\chi\, (l z+ g)=0.
\end{equation}

Using  the Eq.(14) and taking now the time derivative of the Eq.(6) to get
\(\dot f\) and substituting all these into the Eq.(9) we obtain
\begin{equation}
\frac{\ddot G}{G}-\frac{{\buildrel\ldots\over G}}{\ddot G}\, \frac{\dot G}{G}
-K\,\frac{\dot
G}{G}+ \frac{1}{2}\, \dot\Pi^2+\dot\chi^2+\frac{1}{2}\,(\lambda
z+\gamma)^2+(lz+g)^2=0
\end{equation}
where \(K=k^2/4-1/2\).

It follows  from  this equation that the sum of the  last two terms must be
 a constant, therefore \(\lambda=l=0\), which in its turn leads, using the
Eq.(15) to the following relation:
\begin{equation}
\dot\Pi=-\frac{2g}{\gamma}\, \dot\chi.
\end{equation}

The last step before getting to the final equation is to substitute the form
of function \(p=\Pi(t)+\gamma z\) into the Eq.(7) which gives
\begin{equation}
\dot\Pi=\frac{a}{G}
\end{equation}
where \(a \) is a constant.

Finally substituting the Eqs.(17) and (18)  into the Eq.(16) we obtain a
single non linear equation for the evolution of the function \(G\)
\begin{equation}
G\,\ddot G ^2 - {\buildrel\ldots\over G}\, \dot G\, G -  K\, \ddot G\, \dot G^2
+ M^2\, \ddot G +
A^2 \,G^2\, \ddot G  = 0\,,
\end{equation}
where we have introduced the following constants
\begin{equation}
M^2=\frac{a^2}{2}+\frac{\gamma^2 a^2}{4g^2},\quad A^2=\frac{\gamma^2}{2}+g^2.
\end{equation}

To summarize up to here, the Eq.(19) provides a key to solve the
Einstein equations. Once the function \(G\) is found the rest of the
functions describing the geometry and the matter are obtained by the
following expressions:
\begin{eqnarray}
p&=&a\int \frac{dt}{G(t)} + \gamma z\nonumber \\
\phi &=&-\frac{k}{2} \log G -\frac{\gamma
a}{2g}\int\frac{dt}{G(t)}+gz\nonumber\\
f&=&-k\phi +\log \frac{\ddot G}{G}-\log 2\Lambda
\end{eqnarray}
The function \(f\) is derived from the Eq.(6).

The Eq.(19) is very similar to that for the Bianchi I case studied in the
previous paper (Eq.(23) of I). The only difference being in the last term non
linear in \(G\). While in principle one may reduce the order of
this equation it leads to  no simplification since we could not
find a first integral like in the homogeneous case. This complicates
somewhat the search for exact solutions for, one can not integrate
 this equation in general. Yet, one may find some particular solutions to
this equation and to study their behaviour.

Before describing some particular cases we should note that all
the exact solutions described in paper I remain solutions  of the
Eq.(19) when the inhomogeneity term vanishes.

The inhomogeneity of the space time is influenced by both  the scalar
field inhomogeneous mode related to the constant \(g\) and pure
gravitational inhomogeneity coming from the transversal degree of the
gravitational field and related to the constant \(\gamma\). These
terms  act on the dynamics of the transitivity surface area given by
the function \(G\) precisely through the last non linear term of the
Eq.(19).

\section{Explicit Exact Solutions}

As we have mentioned previously all the models given in Paper I
can be considered as particular solutions of the Eqs.(19)-(21) and therefore we
will not return to them here.

By inspection one may see that
\begin{equation}
G=e^t
\end{equation}
and
\begin{equation}
G=\sinhJ\omega t
\end{equation}
both are solutions of the Eq.(19) for particular values of the constants. We
will now look at these solutions separately.

\subsection{\(G=e^t\)}

For this case we obtain for the metric functions and the scalar field the
following expressions:
\begin{eqnarray}
p&=&\gamma z\nonumber\\
\phi&=& -\frac{k}{2} t+ g z\nonumber\\
f&=&\frac{k^2}{2} t-g k z-\log 2\Lambda
\end{eqnarray}
where  the constants are related by
\begin{equation}
2\,\gamma^2+4\,g^2=k^2-2,\quad a=0.
\end{equation}
Note that \(k^2\geq 2\).

To see whether the models defined by the Eqs. (24) and (25) inflate at some
stage
of their history we will define a 4-velocity field \(u^\alpha\) normal to the
hypersurfaces \(\phi=constant\)
\begin{equation}
u^\alpha = \frac{\phi_{,\alpha}}{\sqrt{-\phi_{,\gamma} \phi^{,\gamma}}}\; .
\end{equation}
Note that for the space-times in question the inequality \(-\phi_{,\gamma}
\phi^{,\gamma}\geq 0\) always holds.

We may now compute  the expansion  \(\Theta\)
and the deceleration parameter \(q\).
After some algebra one gets
\begin{equation}
\Theta= \sqrt{2\Lambda}\, \frac{k}{4}\,
\frac{3+\gamma}{\sqrt{\frac{1+\gamma}{2}}}\, e^{-\frac{k^2}{4}t} e^{\frac{g
k }{2} z} \end{equation}
together with \(q\)
\begin{equation}
q=\gamma^2\, \frac{\gamma^2+3}{\gamma^2 +1}\, \frac{k^2}{4}\,
2\Lambda\, \Theta^2\, e^{-\frac{k^2}{2} t} e^{g k  z}\; .
\end{equation}
It is easy to see from the Eq. (28) that \(q\geq 0\). Hence,
these solutions do not undergo inflationary phase, but for \(q=0\)
(\(\gamma=0\)) which defines a kind of a ``weak'' inflation. When \(k^2=2\),
(\(\gamma=g=0\)) one obtains
\begin{eqnarray}
ds^2 & = & \frac{1}{2\Lambda} e^t (-dt^2+dz^2) +e^t(dx^2+dy^2) \nonumber \\
\phi & = & -\frac{1}{\sqrt{2}} t
\end{eqnarray}
Transforming the solution into synchronous coordinates (\(t=2 \log T\)) one
obtains
\begin{eqnarray}
ds^2 & = & -dT^2+ T^2 (dx^2+dy^2+dz^2) \nonumber \\
\phi & = & -\sqrt{2}\, \log T
\end{eqnarray}
representing an isotropic and homogeneous FRW solution first obtained by
Ellis and Madsen \cite{e-m}.

One may look at these solutions yet from a different point of view. Choosing
the 4-velocity as given by the Eq. (26), one may show that the
scalar field stress-energy tensor  takes the perfect fluid
form:
\begin{equation}
T_{\mu\nu} = (p + \rho) u_{\mu} u_\nu + p g_{\mu\nu}.
\end{equation}
with the density and the presure given by
\begin{eqnarray}
\rho & = & -\frac{1}{2} \phi_{,\gamma} \phi^{,\gamma} + V(\phi) \nonumber \\
p & = & -\frac{1}{2} \phi_{,\gamma} \phi^{,\gamma} - V(\phi)
\end{eqnarray}

Substituting \(V\), \(\phi\) and \(u^\alpha\) into these expressions we readily
find that the fluid has a
simple adiabatic equation of state \(p=n\rho\), where
\(n=\frac{\gamma^2-1}{\gamma^2+3}\) (\(-\frac{1}{3}\leq n\leq 1\)) and
\begin{equation}
\rho=\Lambda\, \left(3+\gamma^2\right)\, e^{-\frac{k^2}{2} t}
e^{ kgz}\;.
\end{equation}

While at first sight this solution looks inhomogeneous, in fact it is not and
after some coordinate transformations, using the scalar field as a new time
coordinate, the
line element may be transformed into an explicit Bianchi type VI form.

\subsection{\(G=\sinh\omega t\)}

For this case we have the following relations between the constants:
\begin{eqnarray}
g^2&=&\frac{\omega^2}{4} (k^2-2)-\frac{\gamma^2}{2}\nonumber\\
a^2&=&2g^2\frac{k^2+2}{k^2-2}
\end{eqnarray}
The rest of the metric functions and the scalar field are given by
\begin{eqnarray}
p&=& \frac{a}{\omega} \log \left(\tanh \frac{\omega t}{2}\right) + \gamma
z\nonumber\\
\phi&=&-\frac{k}{2}\log \sinh \omega t -\frac{\gamma}{2\omega}
\sqrt{2\frac{k^2+2}{k^2-2}} \log \left(\tanh \frac{\omega t}{2}\right) +g
z\nonumber\\
f&=&-k\phi+\log\frac{\omega^2}{2\Lambda}
\end{eqnarray}
Let us note again that \(k^2>2\).

To see whether these solutions inflate  one has to proceed as in the
previous model. Technically, however, the expressions in this case start to be
quite long so we shall consider a particular representative case of this
solution by choosing particular values of the free constants. We have chosen
\begin{equation}
\gamma=\frac{k^2}{2},\quad \omega^2=\frac{k^2}{2}\frac{k^2+2}{k^2-2}.
\end{equation}
After some lenghty calculations we get that the sign of \(q\) is determined by
the sign of the following polynomial:
\begin{equation}
\sum_{n=0}^{n=4} c_n \cosh^n \omega t
\end{equation}
where the coefficients \(c_n\) are messy functions of the parameter \(k\). One
may show that for any \(k\) all the coefficients
\(c_n\) are strictly positive. We therefore conclude that this representative
solution never inflates.

We may have as well a  look at the asymptotic behaviour
of the above model at \(t \rightarrow \infty\). It is easy to see that in
general
these models tend to a homogeneous (but anisotropic) universes of Bianchi
type VI. For large values of the parameter \(k\) the solutions approach also
the
Bianchi type VI anisotropic models whereas for values of \(k\) close to
\(\sqrt{2}\) the metric tends to that of Bianchi type III.  After a
coordinate transformation the metric can be cast for \(k\sim\sqrt{2}\) into
the following form:
\begin{equation}
ds^2 \sim -dT^2 + T^2 ( e^{z}dx^2 + dy^2 + dz^2) \; .
\end{equation}

It would be interesting of course to get a general solution of the Eq.(19) as
we did in the Bianchi I case. We are  afraid, however, that technically this
task may turn out to be very difficult. One may probably look for more
particular
solutions of the Eq.(19) yet we feel at this stage that one may proceed to
study the evolution of these models numerically since we have got enough
analytic
exact solutions against which the numerical results can be tested.

\section{The late time behaviour of the generic solution}

In this Section we present the results of the qualitative and numerical
\cite{ode} study  of
the asymptotic behaviour of the generic solution described by the line element
(2).

\subsection{The Qualitative Analysis}

Before presenting the results of the numerical analysis of the evolution
equation (19), we can try to apply to it a qualitative analysis similar to that
of paper I. As we will see in the following, the results will be rather
different.

By using
\begin{equation}
x\,=\, \log G, \qquad y=\dot G e^{(K-1)x},
\end{equation}
the Eq.~(19) reduces to
\begin{equation}
y''+(1-K)\,y'\,=\,e^{2Kx}\left( A^2+e^{-2x} M^2\right)\, \frac{y'+(1-K)y}{y^2}.
\end{equation}
For \(-1/2<K<0\,(k^2<2)\) the right hand side of (40) vanishes when
\(y\rightarrow  E\) and \(x\rightarrow \infty\). So, one can expect the same
asymptotic behaviour as in the homogeneous case: \(G\sim (Ct+D)^{1/K}\). Our
numerical experiments have shown that this is the case.

However, from the Eq.~(40) we see that one can not expect the
same behaviour when \(0<K<1\,(2<k^2<6)\).

In terms of the new variables
\begin{equation}
x\,=\,\log G,\qquad y\,=\,\dot G,
\end{equation}
the Eq.~(19) reduces to the following equation
\begin{equation}
y''+(K-1)\,y'\,=\,\left(M^2+A^2 e^{2x}\right)\, \frac{y'}{y^2}.
\end{equation}
If \(K>1\) (i.e., if \(k^2>6\)), the right hand side of the Eq.~(42) does not
vanish when \(y\rightarrow C\) and \(x\rightarrow \infty\). Therefore one does
not
expect, in this case, that the asymptotic behaviour is of the form \(G\sim
Ct+D\).

To analyze the cases \(K>0\), (\(k^2>2\)), in which we should expect an
asymptotic behaviour different from that of the homogeneous case, let us
consider the following variables:
\begin{equation}
u\,=\,\frac{\dot G}{G},\qquad v\,=\,\frac{\ddot G}{\dot G}\,.
\end{equation}
The evolution equation (19) can be written in the form of a non-linear first
order system:
\begin{eqnarray}
\dot G & = & G\, u, \nonumber \\
\dot u & = & u\,(v-u),  \\
\dot v & = & -K \, u\, v +\left(A^2+\frac{M^2}{G^2}\right)\,\frac{v}{u}\, .
\nonumber
\end{eqnarray}
If \(M=0\), the last two equations form an autonomous system:
\begin{eqnarray}
\dot u &=& u\,(v-u), \nonumber \\
\dot v &=& -K\, u\, v\, + A^2\,\frac{v}{u}\, ,
\end{eqnarray}
which for \(K>0\) has a single equilibrium point \(u=v=A/\sqrt{K}\).
Furthermore, the
characteristic exponents of this equilibrium point are \(-A(1 \pm
\sqrt{1-8K})/2\sqrt{K}\)
and their real parts are always negative. Consequently, the equilibrium point
is
asymptotically stable. This attractor corresponds to the solutions of the form
\(G\propto
\exp (At/\sqrt{K})\). The  corresponding phase-space is depicted in the Fig.~1
for \(K=1/2\) and
\(A=1/2\).

When \(M\neq 0\) there is an additional term, \(M^2v/G^2u\), but we see from
the Eq.~(44)
that it will decrease exponentially as (\(u,v\)) approaches the equilibrium
point. One
thus   expects the same asymptotic behaviour even in this case. For instance,
we can
see in the Fig.~2 the same case as in the Fig.~1 but with \(M=0.001\) (the same
behaviour
is observed for larger values of \(M\)). The solutions corresponding to the
same initial
conditions with \(M=0\) are displayed as dotted lines. As expected, we see that
both cases
are rather different for small values of \(t\) (which correspond to small
\(G\)), but
tend asymptotically to the equilibrium point. Note, that some lines  appear to
cross
because with \(M\neq 0\) the plane (\(u,v\)) is a projection of the three
dimensional
phase-space (\(G,u,v\)). As described in the next subsection, we have found
this behaviour
in all numerical experiments.

\subsection{The Numerical Analysis}

As in the case of homogeneous Bianchi I models (paper I) we look again for
the following asymptotic behaviour suggested by the exact solutions:
\begin{equation}
G\sim t^N
\end{equation}
and
\begin{equation}
G\sim e^{N t}
\end{equation}
which correspond to FRW (Kasner \(N=1\)) or anisotropic behaviour when  the
inhomogeneity is switched off. All the numerical solutions we ever obtained had
one of the
aforementioned asymptotic behaviour. As in paper I we use the following
quantities
\begin{equation} n_{1}(t) =\frac{\dot G^2}{\dot G^2 -G \ddot G}\,,
\end{equation}
and
\begin{equation}
n_{2}(t) =\frac{\ddot G}{\dot G}\,.
\end{equation}
which tend to a constant if each of the asymptotic behaviours occurs. The
Eq.~(48) monitors
the behaviour given by the Eq.~(46), while the Eq.~(49) monitors that one
described by the
Eq.~(47).

We have integrated numerically the Eq.(19), and summarize our results in the
following table
\[
\begin{array}{|c||c|c|c|}
\hline
        & 0<k^2<2 & 2<k^2<6 & k^2>6 \\
\hline\hline
A^2=0 &  &  &  \\
 \hbox{Homogeneous case}   &  G\sim (Ct+D)^{1/K}  &  G\sim (Ct+D)^{1/K} &
G\sim Ct+D\\ \hline
A^2\neq 0 &  &  &  \\
 \hbox{Inhomogeneous case} &  G\sim (Ct+D)^{1/K}  &  G\sim C
e^{At/\sqrt{K}}+D &
  G\sim C e^{At/\sqrt{K}}+D \\ \hline
\end{array}
\]
where \(C\) and \(D\) are constants.

Different initial conditions were used during the numerical integration. We
have
only kept the initial condition \(G(0)=0\) fixed. During the numerical
integration we have always monitored the positivity of the function \(G\) and
of
its second derivative.

Looking at the above table we see that a new type of the asymptotic behaviour
appears  when the inhomogeneity is introduced: \(G\sim e^{At/\sqrt{K}}+D\).
Note, as pointed in a our previous paper, that this type of the behaviour was
structuraly unstable. Surprisingly, the inhomogeneity stabilizes this
asymptotic
solution. The late time exponential behaviour occurs for \(k^2>2\). In this
case
the models tend to those of Bianchi type VI as described in Section 3.2. These
models never
become isotropic, although the universe homogenizes. For \(k^2<2\) the function
\(G\)
tends to \((Ct+D)^{1/K}\) which in absence of inhomogeneity would have been
that of FRW
case. Yet, the inhomogeneity imprints in other metric functions \(p\) and \(f\)
prevent these models to homogenize, in the sense that they do not tend to a
Bianchi model,
let alone to isotropize.

We have also studied numerically the ocurrence of inflation by computing the
sign of the deceleration parameter \(q\). We have noticed that inflation always
occurs for \(k^2<2\) although it takes some time for the model to start
inflating.
For \(k^2>2\) most of the models do not inflate as long as the gradient of the
scalar field
remains timelike. We insist on this condition since otherwise the fluid
interpretation of
the matter field is problematic \cite {T}.

We also find that the introduction of the inhomogeneity may introduce multiple
inflation:
the model starts decelerating then acelerates, then decelarates and acelerates
again. This
never happens for Bianchi I models.

\section{Conclusions}

We have discussed in this paper the simplest inhomogeneous generalizations of
the Bianchi
type I cosmological models with an exponential-potential scalar field.
Restricting the
geometry to be as close as possible to that of Bianchi type I anisotropic
cosmological
model by keeping the element of the transitivity area time-dependent only and
thus
globally timelike, we have been able to reduce the Einstein Equations to a
single
non-linear differential equation. Several exact solutions to this equation, and
cosequently to the full set of the  equations, were presented and discussed.
These
solutions then served us as a bench test to analyse numerically  the
central equation (19)  and the dynamics of the cosmological models.

It is needless to say that the numerical integrations were at each stage tested
against
the analytic results obtained  both for homogeneous and inhomogeneous cases.

For the models we have studied our results are as follows :

a) {\em Homogeneous Anisotropic Case}  ( See as well paper I )
\\[0.2cm]
1. The slope of the potential given by the constant \(k\) is the key factor
influencing
the occurence of inflation and late-time isotropization of the model.
\newline 2. For \(k^2 < 2\)  the models always inflate and isotropize. For \(2
< k^2 <6\) the
models still isotropize, however do not inflate in most of the cases. Yet, the
\(G\sim
t^{1/K}\) (FRW-type) behaviour can not be called an attractor in a strict
technical sense,
since  \(G\sim t^{1/K}\) becomes an exact solution of the Einstein Equations
only when the
integration constants are severely restricted.
\newline 3. For \(k^2 > 6\) the models do not isotropize and have a Kasner-like
asymptotic
behaviour. This is an atractor solution, for, it is a solution of Einstein
Equations with
arbitrary integration constants.

b) {\em The Inhomogeneous case}
\\[0.2cm]
1. The slope of the potential is still of  key importance in the behaviour of
the
cosmological models, in this case, however, the inhomogeneity influences
strongly the
evolution and enriches the behaviour of the models.
\newline 2. The introduction of the inhomogeneity stabilizes the \(G\sim
e^{t/\sqrt{K}}\)
asymptotic behaviour leading always, for \(k^2 > 2\), to an anisotropic Bianchi
type VI
universe.
\newline 3. The solutions \(G\sim t^{1/K} \) which are asymptotically generic
in the case
\(k^2 < 2\) are of no help for isotropization, in this case, for, unlike in the
homogeneous
case, these are not FRW solutions anymore. The spatial dependence of other
metric
functions prevents the homogenization.
\newline 4. As to the inflation , we have found that for \(k^2 < 2\)  the
models do
generically inflate. We have also observed that the introduction of the
inhomogeneity
induces a new type of the dynamical behaviour, not present in Bianchi I models,
the multiple inflation, in which the deceleration parameter \(q\)  changes its
sign several times during the entire history of the universe. The multiple
inflation is, however, a subject to a fine tuning of the integration constants.

We feel that the best way to close is to call for  more work on inhomogeneous
inflation.
To treat the generic inhomogeneous model one certainly needs to use the
numerical
analysis. We hope then that some of our results may be of use for further
numerical
studies of generic inhomogeneous models.
\\[1cm]

This work was supported by the CICYT grant PS90-0093.

\newpage
\centerline{\bf Figure Captions}

\\[0.5cm]
Fig.1~~~~Log-Log plot of the phase space of the Eq.~(45) for \(K=A=1/2\).

\\[0.5cm]
Fig.2~~~~Some solutions of the Eq.~(44) for \(K=A=1/2\) and \(M=0.001\). The
solutions
corresponding to the same initial conditions but \(M=0\) appear as dotted
lines. All
solutions decay to the equilibrium point of the Fig.~1 (Note that the scale has
been
changed.)

\end{document}